\documentclass[11pt]{article}
\usepackage{graphicx}
\topmargin - 2cm
\oddsidemargin - 0.1cm
\evensidemargin - 5cm
\newcommand{\eproof}{\rule{0.2cm}{0.2cm}}

\newtheorem{thm}{Theorem}[section]

\newtheorem{lem}[thm]{Lemma}
\newtheorem{cor}[thm]{Corollary}
\newtheorem{definition}[thm]{Definition}
\newtheorem{remark}[thm]{Remark}

\begin{document}

\title{\small{{\bf SYMMETRIC $(q,\alpha)$-STABLE DISTRIBUTIONS. \\
PART II: SECOND REPRESENTATION}}}
\author{Sabir Umarov$^{1}$, Constantino Tsallis$^{2,3}$, Murray Gell-Mann$^{3}$ \\and Stanly Steinberg$^{4}$}
\date{}
\maketitle
\begin{center}
$^{1}$ {\it Department of Mathematics, Tufts University,
Medford, MA 02155, USA}\\

$^{2}$ {\it Santa Fe Institute \\
1399 Hyde Park Road, Santa Fe, NM 87501,
USA}\\

$^3$ {\it Centro Brasileiro de Pesquisas Fisicas \\
Xavier Sigaud 150, 22290-180 Rio de Janeiro-RJ, Brazil}

$^4$ {\it Department of Mathematics and Statistics\\
University of New Mexico, Albuquerque, NM 87131, USA}
\end{center}


\begin{abstract}
This paper is a continuation of papers
\cite{UmarovTsallisSteinberg,UmarovTsallisGellmannSteinberg}. In
Part I \cite{UmarovTsallisGellmannSteinberg} a description
(representation) of $(q,\alpha)$-stable distributions based on a
$F_q$-transform was given. Here, in Part II, we present another
description of these distributions. This approach generalizes
results of \cite{UmarovTsallisSteinberg} (which corresponds to
$\alpha=2, \, Q\in [1,3)$) to the whole range of stability  and
nonextensivity parameters  $\alpha \in (0,2]$ and $Q \in [1,3),$
respectively. The present case $\alpha=2$ recovers the $q$-Gaussian
distributions. Similar to what is discussed in
\cite{UmarovTsallisSteinberg}, a triplet $(q^{\ast},q,q_{\ast})$
arises for which the mapping $F_{q^{\ast}}: \, \mathcal{G}_{q}
\rightarrow \mathcal{G}_{q_{\ast}}$ holds. Moreover, by unifying the
two preceding descriptions, further possible extensions are
discussed and some conjectures are formulated.
\end{abstract}

\section{Introduction}
In the paper \cite{UmarovTsallisSteinberg}, and in Part I
\cite{UmarovTsallisGellmannSteinberg} of the current paper, we
discussed a $q$-generalization of the classic central limit theorem
(see also \cite{QT,QT2}) applicable to nonextensive statistical
mechanics, (see \cite{Tsallis1988,Tsallis2005,GellmannTsallis} and
references therein), and $q$-generalization of $\alpha$-stable
L\'evy distributions (see, e.g.,
\cite{KolmogorovGnedenko,MeerschaertScheffler,SamorodnitskyTaqqu,UchaykinZolotarev}).
The obtained generalizations concern random variables with a special
long-range correlations arising in nonextensive statistical
mechanics. In \cite{UmarovTsallisSteinberg} we introduced
$q$-Fourier transform \footnote{The $q$-Fourier transform is
formally defined as $F_q[f](\xi)=\int_{R}f \otimes_q e_q^{ix\xi}dx$
and is a nonlinear operator if $q \neq 1$.} and the function $
z(s)={(1+s)}/{(3-s)} $ to describe attractors of scaling limits of
sums of $q$-independent random variables with a finite
$(2q-1)$-variance \footnote{We required there $1 \le q <2$. Denoting
$Q=2q-1$, it is easy to see that this condition is equivalent to the
finiteness of the $Q$-variance with $1 \le Q<3$.}. This description
was essentially based on the mapping
\begin{equation}
\label{paper1}
F_q: \mathcal{G}_q  \rightarrow \mathcal{G}_{z(q)},
\end{equation}
where $\mathcal{G}_q$ is the set of $q$-Gaussians.

In Part I we introduced the set $\mathcal{L}_{q}(\alpha)$ of
$(q,\alpha)$-stable distributions with associated densities having
asymptotic behavior $f \sim C |x|^{-\phi(q,\alpha)}, \,
|x|\rightarrow \infty,$ where $\phi(q,\alpha)=\frac{\alpha +
1}{1+\alpha (q-1)},$ $0<\alpha<2, \, 1 \le q<2$, and $C>0$ is a
constant. The corresponding random variables have infinite
$Q$-variance ($Q=2q-1$) for all $0<\alpha<2$ and $1 \le Q < 3.$
Their representation was obtained  through the map
\begin{equation}
\label{paper2part1}
F_q: \mathcal{G}_{q^{L}} [2] \rightarrow \mathcal{G}_{q}[\alpha],
\end{equation}
where $\mathcal{G}_{q}[\alpha]= \{b e_q^{-\beta |\xi|^{\alpha}}, \,
b>0,  \, \beta >0 \}$ and
$$
q^{L}=\frac{3+Q\alpha}{\alpha + 1}, \, Q=2q-1.
$$
Recall that the parameters $Q$ and $\alpha$ range in the set $
\mathcal{Q}= \{1 \le Q<3, \, \, 0<\alpha \le 2 \}, $ and in the
framework of this description, the value $\alpha = 2$ was peculiar.
For the convenience of the reader we also reproduce the following
lemma proved in Part I, which plays a key role in the current
description as well.

\begin{lem}
\label{mainlemma} Let $f(x), \, x \in R, $ be a symmetric
probability density function of a given random variable. Further,
let either
\begin{itemize}
\item[(i)] the $(2q-1)$-variance $\sigma_{2q-1}^2 < \infty,$
(associated with $\alpha = 2$),
or
\item[(ii)]
$f(x) \in H_{q,\alpha}$, where $(2q-1,\alpha) \in \mathcal{Q}_2.$
\end{itemize}

Then for the $q$-Fourier transform of $f(x)$ the following asymptotic
relation holds true:
\begin{equation}
\label{asforfi}
F_q[f](\xi)= 1 - \mu_{q,\alpha} |\xi|^{\alpha} + o(|\xi|^{\alpha}), \xi
\rightarrow 0,
\end{equation}
where
\begin{equation}
\mu_{q,\alpha} = \left\{ \begin{array}{ll}
  \vspace{1cm}
          \frac{q}{2} \sigma_{2q-1}^2 \nu_{2q-1} , &
          \mbox{if $\alpha = 2$;} \\
          \   \frac{2^{2-\alpha}(1+\alpha(q-1))}{2-q}  \int_0^{\infty}
\frac{- \, \Psi_q (y)}{y^{\alpha+1}} dy, & \mbox{if $0 < \alpha < 2$,}
  \end{array} \right.
\end{equation}
where $\nu_{2q-1}(f)= \int_{-\infty}^{\infty} [f(x)]^{2q-1}  \, dx$,
and $\Psi_q(y)=\cos_q (2x)-1.$ \footnote{For the definition of
$q$-$\cos$ see
\cite{UmarovTsallisSteinberg,qnivanen,qborges1,qborges}.}
\end{lem}

In Part II we represent second description of $(q,\alpha)$-stable
distributions. In the frame of the new description the value
$\alpha=2$ is no longer peculiar, but in this case we need to
separate the value $q = 1. $\footnote{$q = Q = 1$ leads to the
exponential functions, unlike to $q \ne 1$, which is connected
asymptotically with power law functions.} More precisely, we expand
the result of the paper \cite{UmarovTsallisSteinberg} to the region
$$
\mathcal{Q}=\{(Q,\alpha): 1 \le Q < 3, 0<\alpha \leq 2 \},
$$
generalizing the mapping (\ref{paper1}) in the form
\begin{equation}
\label{paper2part2} F_{\zeta_{\alpha}(q)}: \mathcal{G}_q [\alpha]
\rightarrow \mathcal{G}_{z_{\alpha}(q)}[\alpha], \, 1 \le q < 2, \,
0<\alpha \leq 2,
\end{equation}
where
\[
\zeta_{\alpha}(s) = \frac{\alpha + 2(q-1)}{\alpha} \, \, \mbox{and}
\, \, z_{\alpha}(s)=\frac{\alpha q -(q- 1) }{\alpha  - (q-1).}
\]
Note that if $\alpha = 2$, then $\zeta_{2}(q)=q,$
$z_2(q)=(1+q)/(3-q),$ and $\mathcal{G}_q [2]=\mathcal{G}_q,$
recovering the mapping (\ref{paper1}).

\section{$(q,\alpha)$-stable distributions. Second description}

Let $1 < q < 2$, or equivalently, $1<Q<3,\, \, Q=2q-1$. It follows
from the definition of the $q$-exponential that any density function
$g \in \mathcal{G}_q [\alpha]$ has the asymptotic behavior $g \sim b
\, |x|^{-\alpha \over {q-1}}, \, b>0,$ for large $|x|$.  The set of
all functions with this asymptotic we denote by
$\mathcal{B}_q[\alpha]$. It is readily seen that $\mathcal{G}_q
[\alpha] \subset \mathcal{B}_q[\alpha] $. At the same time, for any
density $f \in \mathcal{B}_q[\alpha]$, there exists a unique density
$g \in \mathcal{G}_q[\alpha]$, such that $f \sim g, \, \, |x|
\rightarrow \infty.$ In this sense the two sets
$\mathcal{G}_q[\alpha]$ and $\mathcal{B}_q[\alpha]$ are
asymptotically equivalent (or {\it asymptotically equal}). Having
this in mind  we write (preferably) $\mathcal{G}_q[\alpha]$ instead
of $\mathcal{B}_q[\alpha].$

\begin{lem}
\label{onetoone}
Let $0 < \alpha \le 2$ be fixed. For arbitrary $q_1$ there exists $q_2$ and a one-to-one mapping $\mathcal{M}_{q_1, q_2}$ such that
\[
\mathcal{M}_{q_1,q_2}: \mathcal{G}_{q_1} [\alpha] \rightarrow \mathcal{G}_{q_2} [2].
\]
\end{lem}
\noindent Obviously, if $\alpha = 2$, then $q_1=q_2$ and
$\mathcal{M}_{q_1,q_2}$ is the identity operator. First we find the
relationship between the three indices $q$, $q^{\ast}$ and
$q_{\ast}$ for which the mapping
\begin{equation}
\label{mapping10}
F_{q^{\ast}}: \mathcal{G}_q [\alpha] \stackrel{(a)}{\rightarrow}
\mathcal{G}_{q_{\ast}}[\alpha],
\end{equation}
where $\stackrel{(a)}{\rightarrow}$ means that the mapping is in the sense of the asymptotic equivalence explained above, holds with $\alpha \in (0,2]$.
The exact meaning of (\ref{mapping10}) is
\[
\mathcal{M}^{-1}_{q_{\ast}, z(q_{\ast})}
F_{q^{\ast}} \mathcal{M}_{q,q^{\ast}}
: \mathcal{G}_q [\alpha] \rightarrow \mathcal{G}_{q_{\ast}}[\alpha].
\]
In the case $\alpha = 2$, as we mentioned above, $\mathcal{M}_{q_1,q_2} = I,$ and the
relationships $q^{\ast} = q$ and $q_{\ast} = \frac{1+q}{3-q}$ were found in
\cite{UmarovTsallisSteinberg}, giving (\ref{paper1}).

\begin{lem}
\label{formula1} Assume $0<\alpha \leq 2$ and let the numbers
$q^{\ast}, \, \, q_{\ast}$ and $q$ be connected with the
relationships
\begin{equation}
\label{qast} q^{\ast} = \frac{\alpha - 2(q-1)}{\alpha} \, \,
\mbox{and} \, \, q_{\ast} = \frac{\alpha q +  (q-1)}{\alpha + (q -
q).}
\end{equation}

Then the mapping (\ref{mapping10}) holds true.
\end{lem}

{\it Proof.} Let $f \in \mathcal{G}_q [\alpha]$, which means that
asymptotically
$f(x) \sim b \, |x|^{- \alpha /(q-1)}, \\ x\rightarrow\infty$ with
some $b>0$. We find the $q^{\ast}$-Gaussian with the same
asymptotics at infinity. For a $q^{\ast}$-Gaussian
$G_{q^{\ast}}(\beta; x), \, \, (\beta>0),$ to be asymptotically
equivalent to $f,$ it is necessary
\[
G_{q^{\ast}}(\beta; x) \sim \frac{C}{|x|^{2 \over {q^{\ast}-1}}}
\sim \frac{b}{|x|^{\alpha \over {q-1}}}, \, \, |x|\rightarrow
\infty,
\]
where $C>0$\footnote{Further on $C$ expresses constants with
possibly different values} and $b>0$ some constants. Hence
\[
q^{\ast}= \frac{\alpha+2(q-1)}{\alpha} = 1 + \frac{2(q-1)}{\alpha}.
\]
Further, it follows from Corollary 2.10 of
\cite{UmarovTsallisSteinberg}, that
\[
F_{q^{\ast}}: \, \, \mathcal{G}_{q^{\ast}}[2] \rightarrow
\mathcal{G}_{q_1} [2],
\]
where
\[
q_1 = \frac{1+q^{\ast}}{3-q^{\ast}} = \frac{\alpha + (q-1)}{\alpha -
(q-1)}.
\]
Further, taking into account the asymptotic equality
\[
G_{q_1}(\beta_1; x) \sim \frac{C}{|x|^{2 \over {q_{1}-1}}} \sim
\frac{C}{|x|^{{\alpha} \over {q_{\ast}-1}} }, \, \, |x| \rightarrow
\infty,
\]
we obtain
\[
q_{\ast}=\frac{\alpha q - (q-1)}{\alpha - (q-1)} = 1 + \frac{\alpha
(q-1)}{\alpha  - (q-1)}.
\]
Thus, the mapping (\ref{mapping10}) holds with $q^{\ast}$ and
$q_{\ast}$ in Equation (\ref{qast}). \eproof
\vspace{.3cm}

Let us now introduce two functions that are important for our
further analysis:
\begin{equation}
z_{\alpha}(s)=\frac{\alpha s - (s-1)}{\alpha - (s-1)} = 1 +
\frac{\alpha(s-1)}{\alpha -(s-1)}, \, \, 0<\alpha \le 2, \, \,  s <
\alpha + 1 \,,
\end{equation}
and
\begin{equation}
\zeta_{\alpha}(s) = \frac{\alpha + 2(s-1)}{\alpha} = 1 +
\frac{2(s-1)}{\alpha}, \, \,  0<\alpha \le 2 \,.
\end{equation}
It can be easily verified that $\zeta_{\alpha}(s)=s$ if $\alpha =
2.$

The inverse, $z_{\alpha}^{-1}(t), \, t \in (1 - \alpha, \infty)$, of the the
first function reads
\begin{equation}
z_{\alpha}^{-1}(t) =
         \frac{\alpha t + (t-1)}{\alpha + (t-1 )} = 1 +
\frac{\alpha(t-1)}{\alpha + (t-1)}\,.
\end{equation}
The function $z(s)$ possess the properties:
$z_{\alpha}(\frac{1}{z_{\alpha}(s)})= {1 \over s}$ and
$z_{\alpha}({1 \over s})={1 \over z_{\alpha}^{-1}(s)}\,.$
If we denote $q_{\alpha, 1}=z_{\alpha}(q)$ and $q_{\alpha,
-1}=z_{\alpha}^{-1}(q),$ then
\begin{equation}
\label{qdual1}
z_{\alpha}({1 \over q_{\alpha, 1}})={1 \over q} \, \, \, \, \,
\mbox{and} \, \, \, \, \, z_{\alpha}({1 \over q})={1 \over q_{\alpha, -1}} \,.
\end{equation}

\begin{cor}
\label{cor01} Let $0<\alpha \le 2$ and $1 \le q < \min
\{2,1+\alpha\}.$ Then the following mapping
\[
F_{\zeta_{\alpha}(q)}:\mathcal{G}_q (\alpha)
\stackrel{(a)}{\rightarrow} \mathcal{G}_{z_{\alpha}(q)}(\alpha),
\]
holds.
\end{cor}
\begin{cor}
There exists the following inverse $q$-Fourier transform
\[
F^{-1}_{\zeta_{\alpha}(q)}: \mathcal{G}_{z_{\alpha}(q)}(\alpha)
\stackrel{(a)}{\rightarrow} \mathcal{G}_q (\alpha),   \, \, \,
q<\min\{2,1+\alpha\},
\]
\end{cor}
Using the above mentioned properties of the function $z_{\alpha}(s)$
we can derive a number of useful formulas for the $q$-Fourier
transforms. For instance, if $q<1$, then we have the mappings
$$
F_{\zeta_{\alpha}(\frac{1}{q_{\alpha,1}})}: \, \, \,
\mathcal{G}_{\frac{1}{q_{\alpha, 1}}} (\alpha) \stackrel{(a)}{\rightarrow}
\mathcal{G}_{\frac{1}{q}}(\alpha),
$$
$$
F_{ \zeta_{\alpha}(\frac{1}{q}) }: \, \, \, \mathcal{G}_{\frac{1}{q}}
(\alpha) \stackrel{(a)}{\rightarrow} \mathcal{G}_{\frac{1}{q_{\alpha,-1}}}(\alpha).
$$
The analogous formulas hold for the inverse $q$-Fourier transforms as well.

\begin{figure}[hpt]
\centering
\includegraphics[width=2.5in]{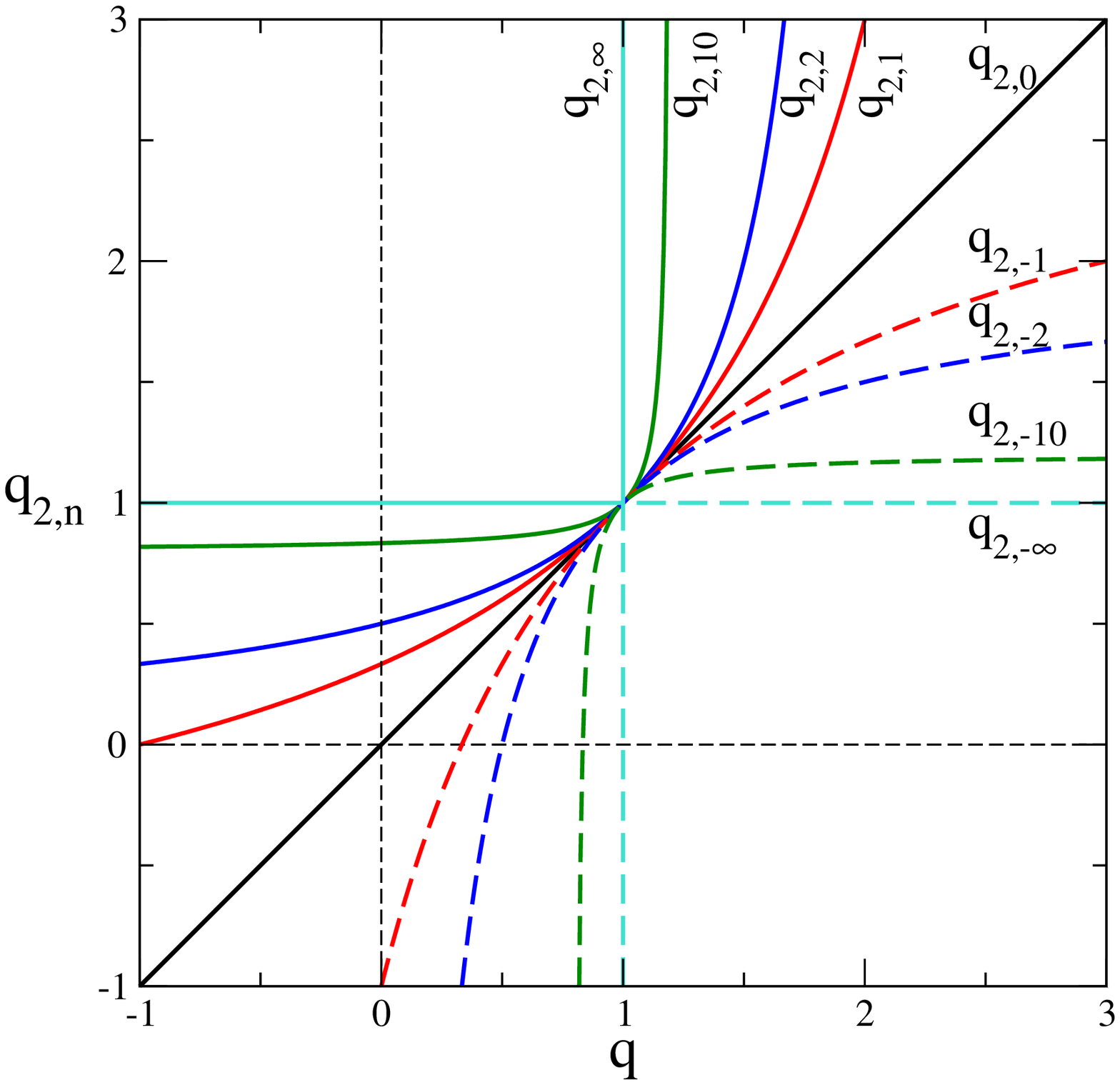}
\hspace{1cm}
\includegraphics[width=2.5in]{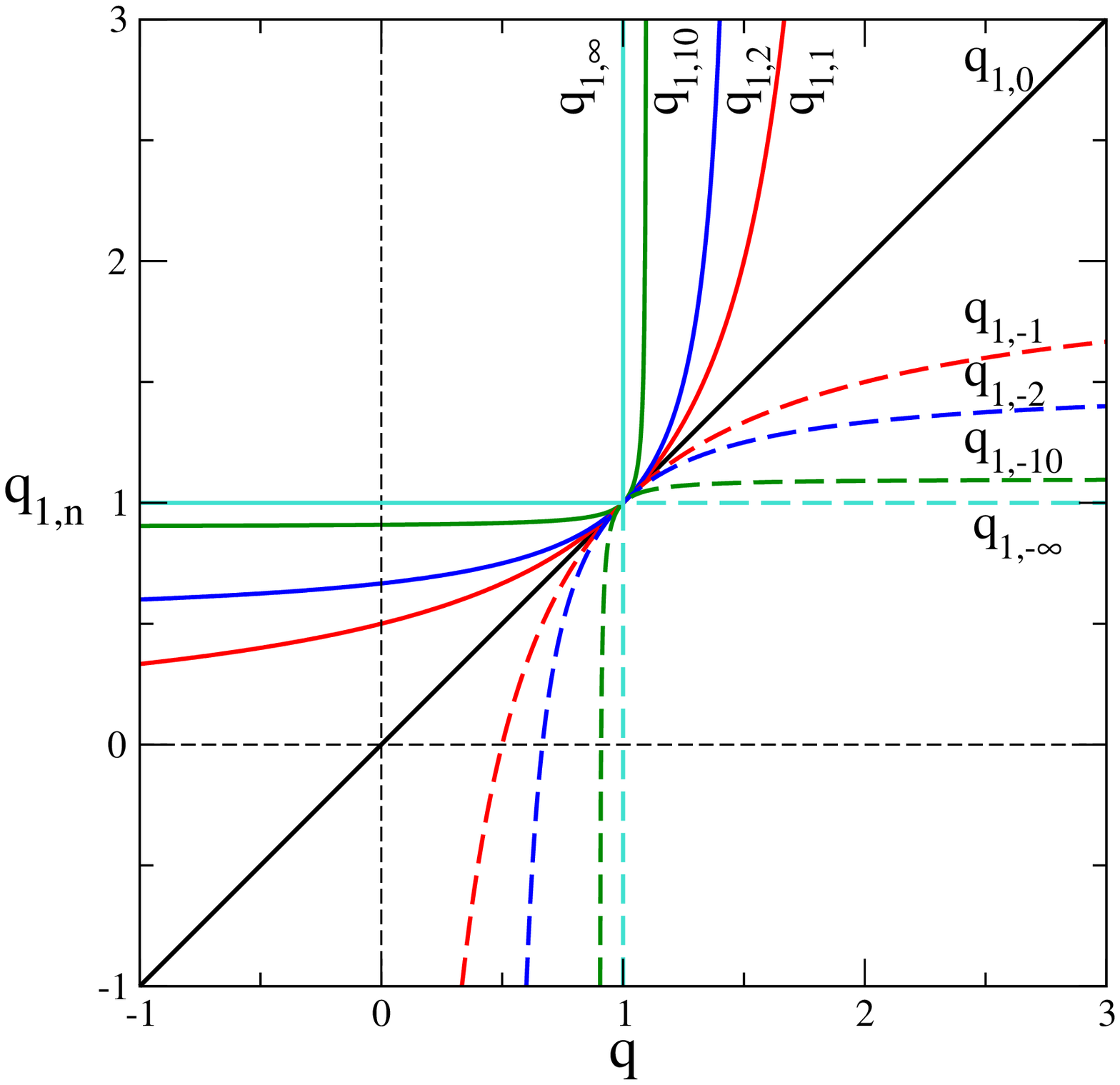}
\caption{\scriptsize The $q$-dependences of $q_{2,n}$ ({\it top})
and $q_{1,n}$ ({\it bottom}) as given by Eq. (12). We notice the
tendency of all the $|n| \to\infty$ curves to collapse onto the
$q_{\alpha,n}=1$ horizontal straight line  if $q \ne 1$, and onto
the $q=1$ vertical straight line if $q=1$. This tendency is
gradually intensified, $\forall n$, when $\alpha$ is fixed onto
values decreasing from 2 towards zero. } \label{fig:qnq}
\end{figure}

Introduce the sequence $q_{\alpha,n} = z_{\alpha,n}(q) =
z(z_{\alpha,n-1}(q)), n=1,2,...,$  with a given $q = z_0(q), \, q<1+\alpha.$
We can extend the sequence $q_{\alpha,n}$ for negative integers
$n=-1,-2,...$ as well putting
$q_{\alpha,-n} = z_{\alpha,-n}(q)=z_{\alpha}^{-1}(z_{\alpha, 1-n}(q)),
n = 1, 2,...\,.$
It is not hard to verify that
\begin{equation}
\label{qn} q_{\alpha, n} = 1 + \frac{\alpha (q-1)}{\alpha - n(q-1)}
= \frac{\alpha q - n(q-1)}{\alpha - n (q-1)}, \, \, -\infty < n \le
[\frac{\alpha}{q-1}].
\end{equation}
The restriction $n\le [\alpha/(q-1)]$ on the right hand side of
(\ref{qn}) comes from intension to have $q_{\alpha,n}>1,$ since
$q$-Fourier transform is defined for $q \ge 1.$ (See Fig.
\ref{fig:qnq} for $q_{\alpha,n}$ with some typical values of
$\alpha$ and $n$).

\vspace{1cm}

Note that $q_{\alpha,n}$ is a function only of $(q,n/\alpha)$, that
$q_{\alpha,n} \equiv 1$ for all $n=0, \pm 1, \pm 2,...,$ if $q=1$,
and that $\lim_{ n \rightarrow \pm \infty}z_{\alpha,n}(q)=1$
for all $q \neq 1.$ Eq. (\ref{qn}) can be rewritten as follows:
\begin{equation}
\frac{\alpha}{q_{\alpha,n}-1}-n=\frac{\alpha}{q-1} \,,      \,\,\,
n=0, \pm1, \pm 2,... \label{rewrite}
\end{equation}
This rewriting puts in evidence an interesting property. If we have a $q$-Gaussian in the variable $|x|^{\alpha/2}$ ($q\ge 1$), i.e., a $q$-exponential in the variable $|x|^\alpha$ (whose asymptotic behavior is proportional to $|x|^{\frac{\alpha}{1-q}}$), its successive derivatives and integrations with regard to $|x|^\alpha$ precisely correspond to $q_{\alpha,n}$-exponentials in the same variable $|x|^\alpha$ (whose asymptotic behavior is proportional to $|x|^{\frac{\alpha}{1-q_{\alpha,n}}}$) \footnote{A typical illustration is as follows. Consider $\alpha=1$, and a normalized $q$-exponential distribution $f(x)$ which identically vanishes for $x<0$, and equals $A(q,\beta)\,e_q^{-\beta x}$ ($1 \le q < 2$, $A(q,\beta)>0$, $\beta>0$) for $x \ge 0$. The {\it accumulated probability} $F(\ge x)=\int_x^{\infty}f(x^\prime)\,dx^\prime$ decreases from unity to zero when $x$ increases from zero to infinity. This probability, frequently appearing in all kinds of applications, is given by $F(\ge x)=\frac{A}{(2-q)\beta} \, [1+(q-1)\beta x]^{-(2-q)/(q-1)}$, i.e., it is proportional to a $q_{1,1}$-exponential with $\frac{1}{1-q_{1,1}}=\frac{1}{1-q} + 1$.
}. Along a similar line, it is also interesting to remark that Eq. (\ref{rewrite}) coincides with Eq. (13) of \cite{MendesTsallis2001} (once we identify the present $\alpha$ with the quantity $z$ therein defined), therein obtained through a quite different approach (related to the renormalization of the index $q$ emerging from summing a specific expression over one degree of freedom).

Let us note also that the definition of the
sequence $q_{\alpha,n}$ (Eq. (\ref{qn})) can be given through the series of mappings
\begin{definition}
\begin{equation}
\label{zright}
... \stackrel{z_{\alpha}}{\rightarrow}q_{\alpha,-2}
\stackrel{z_{\alpha}}{\rightarrow}q_{\alpha,-1}
\stackrel{z_{\alpha}}{\rightarrow}q_{\alpha,0}=q
\stackrel{z_{\alpha}}{\rightarrow}q_{\alpha,1}
\stackrel{z_{\alpha}}{\rightarrow}q_{\alpha,2} \stackrel{z_{\alpha}}{\rightarrow}...
\end{equation}
\begin{equation}
\label{zleft}
...
\stackrel{z_{\alpha}^{-1}}{\leftarrow}q_{\alpha,-2}\stackrel{z_{\alpha}^{-1}}{\leftarrow}q_{\alpha,-1}\stackrel{z_{\alpha}^{-1}}{\leftarrow}q_{\alpha,0}=q
\stackrel{z_{\alpha}^{-1}}{\leftarrow}q_{\alpha,1}
\stackrel{z_{\alpha}^{-1}}{\leftarrow}q_{\alpha,2}
\stackrel{z_{\alpha}^{-1}}{\leftarrow} ...
\end{equation}
\end{definition}
Further, we introduce the sequence
$q^{\ast}_{\alpha,n}=\zeta(q_{\alpha,n})$, which can be written in
the form
\begin{equation}
\label{qstarn} q^{\ast}_{\alpha,n}=1+\frac{2(q-1)}{\alpha - n(q-1)}
= \frac{\alpha + (n-2)(1-q)}{\alpha-n(q-1)}, \, \, n =0,\pm 1,...,
\end{equation}
or, equivalently,
\begin{equation}
\label{qstarn2} \frac{2}{q^{\ast}_{\alpha,n}-1}+n=\frac{\alpha}{q-1}
 \, , \, \, n =0,\pm 1,....
\end{equation}

It follows from Lemma \ref{formula1} and definitions of sequences
$q_{\alpha,n}$ and $q^{\ast}_{\alpha, n}$ that
\begin{equation}
\label{formula2} F_{q^{\ast}_{\alpha,n}}: \, \,
\mathcal{G}_{q_{\alpha,n}}[\alpha] \rightarrow
\mathcal{G}_{q_{\alpha,n+1}}, -\infty < n \le [\frac{\alpha}{q-1}].
\end{equation}
\begin{lem}
\label{dual} For all $n=0, \pm1, \pm2,...$ the following relations
\begin{equation}
\label{dualityrelation}
q^{\ast}_{\alpha, n-1} + \frac{1}{q^{\ast}_{\alpha, n+1}} = 2,
\end{equation}
\begin{equation}
\label{p2n} q^{\ast}_{2,n}=q_{2,n} \, ,
\end{equation}
hold.
\end{lem}
\vspace{.3cm}

{\it Proof.} We notice that
\[
\frac{1}{q^{\ast}_{\alpha, n+1}} = 1 - \frac{2(q-1)}{\alpha -
(n-1)(q-1)}.
\]
By the definition (\ref{qstarn})
\[
-\frac{2(q-1)}{\alpha - (n-1)(q-1)}=1-q_{\alpha,n-1}^{\ast},
\]
 which implies (\ref{dualityrelation})
immediately. The relation (\ref{p2n}) can be checked easily. \eproof
\vspace{.3cm}

\begin{remark}
The property $q^{\ast}_{2,n}=q_{2,n}$ shows that the sequences
(\ref{qn}) and (\ref{qstarn}) coincide if $\alpha=2$. Hence, the mapping
(\ref{formula2}) takes the form
$F_{q_{2,n}}: \, \, \mathcal{G}_{q_{2,n}}(2) \rightarrow
\mathcal{G}_{q_{2,n+1}}(2),$ recovering Lemma 2.16 of
\cite{UmarovTsallisSteinberg}. Moreover,
in this case the relation (\ref{dualityrelation}) holds for the
sequence (\ref{qn}) as well.
If $\alpha < 2$ \footnote{As is known in the classic theory ($q=1$)
this case describes anomalous diffusion processes. If $q=1$, then
$q_{\alpha,n} \equiv q^{\ast}_{\alpha,n} \equiv 1$. In the nonextensive
systems,
as we can see from (\ref{formula2}), there exist two separate
sequences, which characterize the system under study. A physical confirmation of this
theoretical result would be highly interesting.}, then
the values of the sequence (\ref{qstarn}) are splitted  from the values
of $q_{\alpha,n}.$ The shift can be measured as
\[
q_{\alpha,n}-q^{\ast}_{\alpha,n}=\frac{(2-\alpha)(1-q)}{\alpha+n(1-q)},
\]
vanishing for $\alpha=2 \,, \forall q$, or for $q=1 \,, \forall
\alpha$. In the latter case $q_{\alpha,n}=q^{\ast}_{\alpha,n} \equiv
1.$
\end{remark}

Define for $n=0, \pm 1, ..., \, \,  k=1,2,...,$ $n+k \le
[\frac{\alpha}{q-1}]+1,$ the operators
\[
F^k_n (f)=F_{q^{\ast}_{\alpha, n+k-1}} \circ ... \circ F_{q^{\ast}
_{\alpha, n}}[f] =
F_{q^{\ast}_{\alpha, n+k-1}}[...F_{q^{\ast}_{\alpha,
n+1}}[F_{q^{\ast}_{\alpha, n}}[f]]...],
\]
and
\[
F^{-k}_n (f)=F^{-1}_{q^{\ast}_{\alpha, n-k}} \circ ... \circ
F^{-1}_{q^{\ast}_{\alpha, n-1}}[f] =
F^{-1}_{q^{\ast}_{\alpha, n-k}}[...F^{-1}_{q^{\ast}_{\alpha,
n-2}}[F^{-1}_{q^{\ast}_{\alpha, n-1}}[f]]...].
\]
In addition, we assume that $F_{q}^{k}[f]=f,$ if $k=0$ for any
appropriate $q.$

Summarizing the above mentioned
relationships, we obtain the following assertions.
\begin{lem}
The following mappings hold:
\begin{enumerate}
\item
$F_{q^{\ast}_{\alpha,n}}: \, \, \mathcal{G}_{q_{\alpha,n}}(\alpha)
\stackrel{(a)}{\rightarrow} \mathcal{G}_{q_{\alpha,n+1}}(\alpha),
-\infty < n \le [\frac{\alpha}{q-1}];$
\item
$F^k_n:\mathcal{G}_{q_{\alpha,n}} (\alpha)
\stackrel{(a)}{\rightarrow} \mathcal{G}_{q_{\alpha,k+n}} (\alpha),
\, \, \, k=1,2,..., \, \,  n = 0, \pm1,..., -\infty < n+k \le
[\frac{\alpha}{q-1}]+1; $
\item
$\lim_{ k \rightarrow - \infty} F^{k}_n \mathcal{G}_q (\alpha) =
\mathcal{G} (\alpha),  n = 0, \pm 1,...,$
\end{enumerate}
where $\mathcal{G}(\alpha)$ is the set of classic $\alpha$-stable L\'evy
densities.
\end{lem}
\begin{lem}
\label{Fqseries} The following series of mappings hold:
\begin{equation}
\label{Fright}
...  \stackrel{F_{q^{\ast}_{\alpha,-2}}}{\rightarrow}
\mathcal{G}_{q_{\alpha,-1}}(\alpha)
\stackrel{F_{q^{\ast}_{\alpha,-1}}}{\rightarrow} \mathcal{G}_{q}(\alpha)
\stackrel{F_{q^{\ast}_{\alpha, 0}}}{\rightarrow}
\mathcal{G}_{q_{\alpha,1}}(\alpha)
\stackrel{F_{q^{\ast}_{\alpha,1}}}{\rightarrow}
\mathcal{G}_{q_{\alpha,2}}(\alpha)  \stackrel{F_{q^{\ast}_{\alpha,2}}}{\rightarrow} ...
\end{equation}
\begin{equation}
\label{Fleft}
...
\stackrel{F^{-1}_{q^{\ast}_{\alpha,-2}}}{\leftarrow}
\mathcal{G}_{q_{\alpha,-1}} (\alpha)
\stackrel{F^{-1}_{q^{\ast}_{\alpha,-1}}}{\leftarrow}
\mathcal{G}_{q}(\alpha)  \stackrel{F^{-1}_{q^{\ast}_{\alpha,0}}}{\leftarrow}
\mathcal{G}_{q_{\alpha,1}}(\alpha)
\stackrel{F^{-1}_{q^{\ast}_{\alpha,1}}}{\leftarrow}
\mathcal{G}_{q_{\alpha,2}} (\alpha) \stackrel{F^{-1}_{q^{\ast}_{\alpha,2}}}{\leftarrow} ...
\end{equation}
\end{lem}

{\bf Theorem 1.} {\it Assume $0<\alpha \leq 2$ and a sequence
$q_{\alpha,n}, \, -\infty < n \le [\alpha/(q-1)],$ is given as in
(\ref{zright}) with $q_0=q \in [1, \min \{2, 1+\alpha\}  ).$ Let
$X_1,X_2,...,X_N,...$ be symmetric $q_{\alpha, k}$-independent (for
some $-\infty < k \le [\alpha/(q-1)]$ and $\alpha \in (0,2])$)
random variables all having the same probability density function
$f(x)$ satisfying the conditions of Lemma \ref{mainlemma}.

Then the sequence
$$Z_N = \frac{X_1+...+X_N}
{(\mu_{q_{\alpha,k},\alpha}N)^{\frac{1}{\alpha (2-q_{\alpha,k})}}}
\, ,
$$ is $q_{\alpha,k}$-convergent\footnote{The definition of $q$-convergence and its relationship to
weak $q$-convergence, see Part I
\cite{UmarovTsallisGellmannSteinberg}.} to a
$(q_{\alpha,k-1},\alpha)$-stable distribution, as $N \rightarrow
\infty.$
\par
Proof.} The case $\alpha=2$ coincides with Theorem 1 of
\cite{UmarovTsallisSteinberg}. For $k=0$, the first part of Theorem
($q$-convergence) is proved in Part I of the paper. The same method
is applicable for $k \neq 1.$ For the readers convenience we proceed
the proof of the first part also in the general case, namely for
arbitrary $k.$
\par
Assume $0<\alpha<2$. We evaluate $F_{q_{\alpha,k}}(Z_N).$ Denote
$Y_j = X_j/{s_N(q_{\alpha,k})}, j=1,2,...,$ where
${s_N(q_{\alpha,k})}= {(\mu_{q_{\alpha,k},\alpha}N)^{\frac{1}{\alpha
(2-q_{\alpha,k})}}}.$ Then $Z_N = Y_1 +...+Y_N.$ It is not hard to
verify that, for a given random variable $X$ and real $a>0$, the
relationship $F_q [aX](\xi)=F_q[X](a^{2-q} \xi)$, is true for
arbitrary $q \ge 1$. It follows from this equality that
$F_{q_{\alpha,k}}(Y_1)=F_{q_{\alpha,k}}[f]( \frac{\xi}{
(\mu_{q_{\alpha,k},\alpha}N)^{\frac{1}{\alpha}} } ).$ Moreover, it
follows from  $q_{\alpha,k}$-independence of $X_1,X_2,...$ and the
associativity property of the $q$-product that
\begin{equation}
\label{step100}
F_{q_{\alpha,k}}[Z_N](\xi)= F_{q_{\alpha,k}}[f]( \frac{\xi}{
(\mu_{q_{\alpha,k},\alpha}N)^{\frac{1}{\alpha}}  } )
{\otimes_{q_{\alpha,k}} ... \otimes_{q_{\alpha,k}}}
F_{q_{\alpha,k}}[f]( \frac{\xi}{
(\mu_{q_{\alpha,k},\alpha}N)^{\frac{1}{\alpha}}  } ) \,\,(N\,\mbox{factors}).
\end{equation}
Hence, making use of the properties of the $q$-logarithm, from
(\ref{step100})
we obtain
\[
\ln_{q_{\alpha,k}} F_{q_{\alpha,k}}[Z_N](\xi)= N \ln_{q_{\alpha,k}}
F_{q_{\alpha,k}}[f]( \frac{\xi}{
(\mu_{q_{\alpha,k},\alpha}N)^{\frac{1}{\alpha}} } ) =
N \ln_{q_{\alpha,k}} ( 1- \frac{|\xi|^{\alpha}}{N} +
o(\frac{|\xi|^{\alpha}}{N})) =
\]
\begin{equation}
\label{step101}
- |\xi|^{\alpha} + o(1), \, N \rightarrow \infty \,,
\end{equation}
locally uniformly by $\xi$.
\par
Consequently, locally uniformly by $\xi,$
\begin{equation}
\label{step_50}
\lim_{N \rightarrow \infty} F_{q_{\alpha,k}}(Z_N) = e_{q_{\alpha,k}}^{-
|\xi|^{\alpha}} \in \mathcal{G}_{q_{\alpha,k}} (\alpha) \,.
\end{equation}
Thus, $Z_N$ is $q_{\alpha,k}$-convergent.
\par
To show the second part of Theorem we use Lemma \ref{Fqseries}. In
accordance with this lemma there exists a density $f(x) \in
\mathcal{G}_{q_{\alpha,k-1}}[\alpha],$ such that
$F_{q^{\ast}_{\alpha,k-1}}[f] = e_{q_{\alpha,k}}^{-
|\xi|^{\alpha}}.$ Hence, $Z_N$ is $q_{\alpha,k}$-convergent to a
$(q_{\alpha,k-1},\alpha)$-stable distribution, as $N \rightarrow
\infty.$ \eproof

\section{Scaling rate analysis}
In paper \cite{UmarovTsallisSteinberg} we obtained the formula
\begin{equation}
\label{betak}
\beta_k = \Bigl(\frac{3-q_{k-1}}{4 q_k
C_{q_{k-1}}^{2 q_{k-1} -2}}\Bigr)^{1 \over {2-q_{k-1}}}.
\end{equation}
for the $q$-Gaussian parameter $\beta$ of the attractor. It follows
from this formula that the scaling rate in the case $\alpha = 2$ is
\begin{equation}
\delta=\frac{1}{2-q_{k-1}} = q_{k+1},
\end{equation}
where $q_{k-1}$ is the $q$-index of the attractor. Moreover, if we
insert the 'evolution parameter' $t$, then the translation of a
$q$-Gaussian to a density in $\mathcal{G}_q[\alpha]$ changes $t$ to
$t^{2/\alpha}.$ Hence, applying these two facts to the general case,
$0<\alpha \le 2,$ and taking into account that the attractor index
in our case is $q^{\ast}_{\alpha, k-1}$, we obtain the formula for
the scaling rate
\begin{equation}
\delta=\frac{2}{\alpha (2-q^{\ast}_{\alpha, k-1})}.
\end{equation}
In accordance with Lemma \ref{dual},  $2-q^{\ast}_{\alpha, k-1} =
1/q^{\ast}_{\alpha, k+1}.$ Consequently,
\begin{equation}
\label{scaling1} \delta=\frac{2}{\alpha} q^{\ast}_{\alpha, k+1} =
\frac{2}{\alpha} \frac{\alpha - (k-1)(q-1)}{\alpha - (k+1)(q-1)}.
\end{equation}
Finally, in terms of $Q=2q-1$ the formula (\ref{scaling1}) takes the
form
\begin{equation}
\label{scaling2} \delta=\frac{2}{\alpha} \frac{2 \alpha -
(k-1)(Q-1)}{2 \alpha - (k+1)(Q-1)}.
\end{equation}
In \cite{UmarovTsallisSteinberg} we noticed that the non-linear
Fokker-Planck equation corresponds to the case $k=1.$ Taking this
fact into account we can {\it conjecture that the fractional
generalization of the nonlinear Fokker-Planck equation is connected
with the scaling rate $$\delta=\frac{2}{\alpha +1 - Q},$$} which can
be derived from (\ref{scaling2}) putting $k=1$. In the case
$\alpha=2$ we get the known result $\delta = 2/(3-Q)$ obtained in
\cite{TsallisBukman}.

\section{Remark on additive and multiplicative dualities}\label{dualitysection}

In the nonextensive statistical mechanical literature, there are two transformations that appear quite frequently in various contexts. They are sometimes referred to as {\it dualities}.  The {\it multiplicative duality} is defined through
\begin{equation}
\mu (q)=1/q \,,
\end{equation}
and the {\it additive duality} is defined through
\begin{equation}
\nu (q)=2-q \,.
\end{equation}
They satisfy $\mu^2=\nu^2={\bf 1}$, where $\bf 1$ represents the {\it identity}, i.e., ${\bf 1}(q)=q, \forall q$.
We also verify that
\begin{equation}
(\mu\nu)^m(\nu\mu)^{m}= (\nu\mu)^m(\mu\nu)^{m}={\bf 1}\;\;\;\;(m=0,1,2,...) \,.
\end{equation}
Consistently, we define $(\mu\nu)^{-m} \equiv (\nu\mu)^{m}$, and $(\nu\mu)^{-m} \equiv (\mu\nu)^{m}$ .

Also, for $m=0,\pm 1, \pm2, ...$, and $\forall q$,
\begin{equation}
\label{duality}
(\mu\nu)^m(q) = \frac{m-(m-1)\,q}{m+1-m\,q} =\frac{q+m(1-q)}{1+m(1-q)} \,,
\end{equation}
\begin{equation}
\nu(\mu\nu)^m(q) = \frac{m+2-(m+1)\,q}{m+1-m\,q} =\frac{2-q+m(1-q)}{1+m(1-q)} \,,
\end{equation}
and
\begin{equation}
(\mu\nu)^m\mu(q) = \frac{-m+1+m\,q}{-m+(m+1)\,q} =\frac{1-m(1-q)}{q-m(1-q)} \,.
\end{equation}

We can easily verify, from Eqs. (\ref{qn}) and (\ref{duality}), that
the sequences $q_{2,n}$ ($n=0,\pm 2,\pm 4, ...$) and $q_{1,n}$
($n=0,\pm 1,\pm 2, ...$) coincide with the sequence $(\mu\nu)^m(q)$
($m=0, \pm 1,\pm ,2,...$).

\section{Conclusion and conjectures}

The $q$-CLT formulated in \cite{UmarovTsallisSteinberg} states that
an appropriately scaled limit of sums of $q_k$-independent random
variables with a finite $(2q_k-1)$-variance is a
$q^{\ast}_k$-Gaussian, which is the $q^{\ast}_k$-Fourier preimage of
a $q_k$-Gaussian. Here $q_k$ and $q^{\ast}_k$ are sequences defined
as
\[
q_k=\frac{2q-k(q-1)}{2-k(q-1)}, \, k = 0, \pm 1, ...,
\]
and
\[
q_k^{\ast}=q_{k-1}, \, k=0, \pm 1, ....
\]
Schematically this theorem can be represented as
\begin{equation}
\label{schemeCLT} \{f:\sigma_{2q_k-1}(f)<\infty\}
\stackrel{F_{q_k}}{\longrightarrow} \mathcal{G}_{q_k}[2]
\stackrel{F_{q^{\ast}_{k}}}{\longleftarrow}
\mathcal{G}_{q_k^{\ast}}[2]
\end{equation}
where  $\mathcal{G}_{q}[2]$ is the set of $q$-Gaussians. We have
noted that the processes described by the $q$-CLT can be effectively
described by the triplet $(P_{att},P_{cor},P_{scl})$, where
$P_{att}, \, P_{cor}$ and  $ P_{scl}$ are parameters of {\it
attractor}, {\it correlation} and {\it scaling rate}, respectively.
We found that (see details in \cite{UmarovTsallisSteinberg})
\begin{equation}
\label{triplet}
(P_{att},P_{cor},P_{scl}) \equiv (q_{k-1}, q_k, q_{k+1}).
\end{equation}
\par
In Part I of this work we discussed a representation of symmetric
$(q,\alpha)$-stable distributions distributions. Schematically the
corresponding theorem (Theorem 1 of
\cite{UmarovTsallisGellmannSteinberg}) is represented as
\begin{equation}
\label{schemeLevy1}
 \mathcal{L}(q,\alpha)
\stackrel{F_{q}}{\longrightarrow} \mathcal{G}_{q}(\alpha) \stackrel{F_q}{\longleftarrow}  \mathcal{G}_{q^{L}} (2), \, 0<\alpha<2,
\end{equation}
where $\mathcal{L}(q,\alpha)$ is the set of $(q,\alpha)$-stable
distributions, $\mathcal{G}_{q^{L}} (2)$ is the set of $q^L$-Gaussians
asymptotically
equivalent to the densities $f \in \mathcal{L}(q,\alpha).$ The index
$q^L$ is linked with $q$ as follows
\[
q^L = q^{L}_{\alpha}(q) = \frac{3+(2q-1)\alpha}{1+\alpha}.
\]
Note that the case $\alpha=2$ is peculiar and we agree to refer to the scheme (\ref{schemeCLT}) in this case.
\par
In the present paper (Part II, Theorem 1) we have studied a
$q$-generalization of the CLT to the case when the $(2q-1)-$variance
of random variables is infinite.  The theorem that we have obtained
generalizes the $q$-CLT, which corresponds to $\alpha=2$, to the
full range $0<\alpha \le 2$. Schematically this theorem can be
represented as
\begin{equation}
\label{schemeLevy2}
\mathcal{L}(q_{\alpha,k},\alpha)
\stackrel{F_{q_{\alpha,k}}}{\longrightarrow} \mathcal{G}_{q_{\alpha,k}}(\alpha)
\stackrel{F_{q^{\ast}_{\alpha,k}}}{\longleftarrow}
\mathcal{G}_{q_{\alpha,k}^{\ast}}(2), \, 0<\alpha \le 2,
\end{equation}
generalizing the scheme (\ref{schemeCLT}). The sequences
$q_{\alpha,k}$ and $q^{\ast}_{\alpha,k}$ in this case read
\[
q_{\alpha,k}=\frac{\alpha q+k(1-q)}{\alpha+k(1-q)}, \, k = 0, \pm 1,
...,
\]
and
\[
q_{\alpha,k}^{\ast}= 1 - \frac{2(1-q)}{\alpha + k (1-q)}, \, k=0, \pm
1, ....
\]
Note that the triplet $(P_{att},P_{cor},P_{scl})$ mentioned above takes, in
this case, the form
\[
(P_{att},P_{cor},P_{scl}) \equiv (q^{\ast}_{\alpha,k-1}, \,
q_{\alpha,k},\, (2/\alpha) q^{\ast}_{\alpha,k+1}),
\]
which coincides with (\ref{triplet}) if $\alpha = 2.$
\par
Finally, unifying the schemes (\ref{schemeLevy1}) and
(\ref{schemeLevy2}) we obtain the general picture for the
description of $(q,\alpha)$-stable distributions:
\begin{equation}
\label{schemeLevy3} \mathcal{L}[q_{\alpha,k},\alpha]
\stackrel{F_{q_{\alpha,k}}}{\longrightarrow}
\mathcal{G}_{q_{\alpha,k}}[\alpha]
\stackrel{F_{q^{\ast}_{\alpha,k}}}{\longleftrightarrow}
\mathcal{G}_{q_{\alpha,k}^{\ast}}[2]
\end{equation}
\hspace{3.2in} $\updownarrow \, F_{q}$
\[
\hspace{0.1in} \mathcal{G}_{q^{L}_{\alpha,k}} [2],
\]
where
\[
 q^{L}_{\alpha,k} = q^{L}_{\alpha}(q_{\alpha,k}) = \frac{3+(2q_{\alpha, k}-1)\alpha}{1+\alpha} \,.
\]

\begin{figure}[hpt]
\vspace{-1.2cm}
\begin{center}
\includegraphics[width=4.0in]{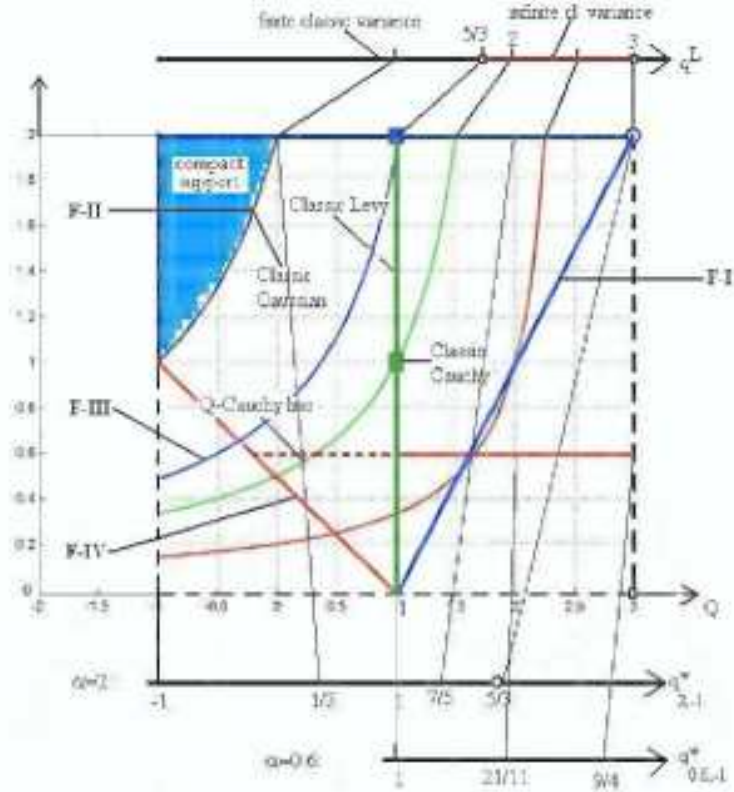}
\end{center}
\vspace{-1.0cm}
\caption{\scriptsize $(Q,\alpha)$-regions (see the
text). } \label{fig:conclusion}
\end{figure}

In Fig. \ref{fig:conclusion} connections of parameters $(Q,\alpha)
\in \mathcal{Q}$ with $q^L$ and $q^{\ast}, \, (k=0)$ is represented.
If $Q = 1$ and $\alpha = 2$ (the blue box in the figure), then the
random variables are independent in the usual sense and have {\it
finite} variance. The standard CLT applies, and the attractors are
classic Gaussians.

If $Q$ belongs to the interval $(1, 3)$ and $\alpha = 2$ (the blue
straight line on the top), the random variables are {\it not}
independent. If the random variables have a {\it finite} Q-variance,
then $q$-CLT \cite{UmarovTsallisSteinberg} applies, and the
attractors belong to the family of $q^{\ast}$-Gaussians. Note that
$q^{\ast}$ runs in $[1, 5/3).$ Thus, in this case, attractors
($q^{\ast}$-Gaussians) have {\it finite} classic variance (i.e.,
$1$-variance) in addition to {\it finite} $q^{\ast}$-variance.

If Q = 1 and $0 < \alpha <2$ (the vertical green line in the
figure), we have the classic L\'evy distributions, and random
variables are independent, and have {\it infinite} variance. Their
scaling limits-attractors belong to the family of $\alpha$-stable
L\'evy distributions. It follows from (20, \, Part I) that in terms
of $q$-Gaussians classic symmetric $\alpha$-stable distributions
correspond to $\cup_{5/3<q<3} \mathcal{G}_q[2].$

If $0 < \alpha < 2,$ and $Q$ belong to the interval $(1, 3)$ we
observe the rich variety of possibilities of $(q,\alpha)$-stable
distributions. In this case random variables are {\it not}
independent, have {\it infinite} variance and {\it infinite
Q-variance}. The rectangle $\{1<Q<3; \, \, 0<\alpha<2 \}$, at the
right of the classic L\'evy line, is covered by non-intersecting
curves
$$
C_{q^L} \equiv \{(Q,\alpha): \frac{3+Q\alpha}{\alpha + 1} = q^L\}, \, 5/3<q^L<3 \,.
$$
In accordance with \cite{UmarovTsallisGellmannSteinberg}, these
families of curves describe all $(Q,\alpha)$-stable distributions
based on the mapping (\ref{schemeLevy1}) with $q$-Fourier transform.
The constant $q^L$ is the index of the $q^L$-Gaussian attractor
corresponding to the points $(Q,\alpha)$ on the curve $C_{q^L}$. For
example, the green curve corresponding to $q^L=2$ describes all
$Q$-Cauchy distributions, recovering the classic Cauchy-Poisson
distribution if $\alpha = 1$ (the green box in the figure). Every
point $(Q,\alpha)$ lying on the brown curve corresponds to
$q^L=2.5$.

The second description of $(Q,\alpha)$-stable distributions
presented in the current paper, and based on the mapping
(\ref{schemeLevy2}) with $q^{\ast}$-Fourier transform leads to a
covering of $\mathcal{Q}$ by curves distinct from $C_{q^L}$. Namely,
in this case we have the following family of straight lines
\begin{equation}
\label{lines} S_{q^{\ast}} \equiv \{(Q,\alpha): \frac{4 \alpha}{Q +
2 \alpha - 1} = 3 - q^{\ast} \}, \, 1 \le q^{\ast} <3,
\end{equation}
which are obtained from (\ref{qstarn}) replacing $n=-1$ and
$2q-1=Q.$ For instance, every $(Q,\alpha)$ on the line F-I (the blue
diagonal of the rectangle in the figure) identifies
$q^{\ast}$-Gaussians with $q^{\ast}=5/3$. This line is the frontier
of points $(Q,\alpha)$ with finite and infinite classic variances.
Namely, all $(Q,\alpha)$ above the line F-I identify attractors with
{\it finite} variance, and points on this line and below identify
attractors with {\it infinite} classic variance. Two bottom lines in
Fig. \ref{fig:conclusion} reflect the sets of $q^{\ast}$
corresponding to lines $\{1\le Q<3; \alpha = 2)\}$ (the top boundary
of the rectangle in the figure) and $\{1\le Q<3; \alpha = 0.6)\}$
(the brown horizontal line in the figure).

\vspace{.2cm}

{\bf Some conjectures.} Both descriptions of $(Q,\alpha)$-stable
distributions are restricted to the region $Q=\{1\le Q<3, 0<\alpha
\le 2\}.$ This limitation is caused by the tool used for these
representations, namely, $Q$-Fourier transform is defined for $Q\ge
1.$ However, at least two facts, the positivity of $\mu_{q,\alpha}$
in Lemma \ref{mainlemma} for $q>\max \{0,1-1/\alpha\}$ (or, the
same, $Q>\max \{-1, 1-2/\alpha\}$) and continuous extensions of
curves in the family $C_{q^L}$, strongly indicate to following
conjectures, regarding the region on the left to the vertical green
line (the classic L\'evy line) in Fig. \ref{fig:conclusion}. In this
region we see three frontier lines, F-II, F-III and F-IV.

{\it Conjecture 1.} The line F-II splits the regions where the
random variables have {\it finite} and {\it infinite} $Q$-variances.
More precisely, the random variables corresponding to $(Q,\alpha)$
on and above the line F-II have a {\it finite} $Q$-variance, and,
consequently, $q$-CLT \cite{UmarovTsallisSteinberg} applies.
Moreover, as seen in the figure, the $q^L$-attractors corresponding
to the points on the line F-II are the classic Gaussians, because
$q^L=1$ for these $(Q,\alpha)$. It follows from this fact, that
$q^L$-Gaussians corresponding to points above F-II have compact
support (the blue region in the figure), and $q^L$-Gaussians
corresponding to points on this line and below have infinite
support.

{\it Conjecture 2.} The line F-III splits the points $(Q,\alpha)$
whose $q^L$-attractors have {\it finite} or {\it infinite} classic
variances. More precisely, the points $(Q,\alpha)$ above this line
identify attractors (in terms of $q^L$-Gaussians) with {\it finite}
classic variance, and the points on this line and below identify
attractors with {\it infinite} classic variance.

{\it Conjecture 3.} The frontier line F-IV with the equation $Q + 2
\alpha -1 = 0$ and joining the points $(1,0)$ and $(-1,1)$ is
related to attractors in terms of $q^{\ast}$-Gaussians. It follows
from (\ref{lines}) that for $(Q,\alpha)$ lying on the line F-IV, the
index $q^{\ast}=-\infty$. Thus the horizontal lines corresponding to
$\alpha < 1$ can be continued only up to the line F-IV with
$q^{\ast} \in (-\infty, 3-\frac{4 \alpha}{Q + 2 \alpha - 1})$ (see
the dashed horizontal brown line in the figure). If $\alpha
\rightarrow 0,$ the $Q$-interval becomes narrower, but
$q^{\ast}$-interval becomes larger tending to $(-\infty, 3)$.

Results confirming or refuting any of these conjectures would be an
essential contribution to the understanding of the nature of
$(Q,\alpha)$-stable distributions, and nonextensive statistical
mechanics, in particular.

\vspace{.2cm}

Let us stress that Fig. \ref{fig:conclusion} corresponds to the
case $k=0$ in the description (\ref{schemeLevy3}). The cases $k \neq
0$ can be analyzed in the same way.

The remarks made in section \ref{dualitysection} 
establish a remarkable connection between sequences which
emerge naturally within the context of the $q$-generalized central
limit theorems, and the elementary dualities that we introduced
in the present work. However, its physical interpretation is yet to be found. It
might be especially interesting if we take into account the fact
that such a connection could be a crucial step (see footnote of page
15378 in \cite{TsallisGellmannSato}) for understanding  the
$q$-triplet that was observed by NASA  using data received from the
spacecraft Voyager 1 . Indeed, the existence of a $q$-triplet,
namely $(q_{sen},q_{rel},q_{stat})$, related respectively to
sensitivity to the initial conditions, relaxation, and stationary
state) was conjectured in \cite{Tsallistriplet}, and was observed in
the solar wind at the distant heliosphere
\cite{BurlagaVinas,BurlagaNessAcuna}.

Finally, let us mention that Parts I and II of the present work respectively correspond, for fixed $(q,\alpha)$, to the {\it distant} and {\it intermediate} regions of Table 1 and Fig. 4 of \cite{QT2}.

\subsection*{Acknowledgments}

We acknowledge thoughtful remarks by R.
Hersh, E.P. Borges and S.M.D. Queiros.
Financial support by the Fullbright Foundation, SI International,
AFRL and NIH grant P20 GMO67594 (USA agencies), and CNPq, Pronex and Faperj (Brazilian agencies) are acknowledged as well.


\begin{thebibliography}{9999}

\bibitem{UmarovTsallisSteinberg}
S. Umarov, C. Tsallis and S. Steinberg, {\it On a $q$-central limit theorem consistent with nonextensive statistical mechanics}, Milan J. Math. {\bf 76} (2008) [DOI 10.1007/s00032-008-0087-y].

\bibitem{UmarovTsallisGellmannSteinberg}
S. Umarov, C. Tsallis, M. Gell-Mann  and S. Steinberg, {\it Symmetric $(q,\alpha)$-stable distributions. Part I: First representation}, preprint (2008).

\bibitem{QT}C. Tsallis and S.M.D. Queiros, {\it Nonextensive statistical mechanics and central limit theorems I - Convolution of independent random variables and $q$-product}, in  {\it Complexity, Metastability and Nonextensivity},  eds. S. Abe, H.J. Herrmann, P. Quarati, A. Rapisarda and C. Tsallis, American Institute of Physics Conference Proceedings {\bf 965}, 8-20 (New York, 2007).

\bibitem{QT2}S.M.D. Queiros and C. Tsallis, {\it Nonextensive statistical mechanics and central limit theorems II - Convolution of     $q$-independent random variables}, in  {\it Complexity, Metastability and Nonextensivity},  eds. S. Abe, H.J. Herrmann, P. Quarati, A. Rapisarda and C. Tsallis, American Institute of Physics Conference Proceedings {\bf 965}, 21-33 (New York, 2007).

\bibitem{Tsallis1988}C. Tsallis, {\it Possible generalization of
Boltzmann-Gibbs statistics},  J. Stat. Phys. {\bf 52}, 479 (1988).
See also E.M.F. Curado and C. Tsallis, {\it Generalized statistical
mechanics: connection with thermodynamics}, J. Phys. A {\bf 24}, L69
(1991) [Corrigenda: {\bf 24}, 3187 (1991) and {\bf 25}, 1019
(1992)], and C. Tsallis, R.S. Mendes and A.R. Plastino, {\it The
role of constraints within generalized nonextensive statistics},
Physica A {\bf 261}, 534 (1998).

\bibitem{Tsallis2005}C. Tsallis, {\it Nonextensive statistical
mechanics, anomalous diffusion and central limit theorems}, Milan
Journal of Mathematics {\bf 73}, 145  (2005).

\bibitem{GellmannTsallis}M. Gell-Mann and C. Tsallis, {\it Nonextensive
Entropy - Interdisciplinary Applications} (Oxford University Press,
New York, 2004).

\bibitem{KolmogorovGnedenko}{B.V. Gnedenko, A.N. Kolmogorov}, {\it Limit Distributions for Sums of Independent Random Variables,} 1954, Addison-Wesley, Reading.

\bibitem{MeerschaertScheffler}{M.M. Meerschaert, H.-P. Scheffler}, {\it Limit Distributions for Sums of Independent Random Vectors. Heavy Tails in Theory and Practice}, {John Wiley and Sons, Inc}, {2001}.

\bibitem{SamorodnitskyTaqqu}{G. Samorodnitsky and M.S. Taqqu}, {\it Stable non-Gaussian Random Processes}, {Chapman and Hall }, {New York}, {1994}.

\bibitem{UchaykinZolotarev}{V.V. Uchaykin and V.M. Zolotarev}, {\it Chance and Stability. Stable Distributions and their Applications}, {VSP}, {Utrecht}, {1999}.

\bibitem{qnivanen}L. Nivanen, A. Le Mehaute and Q.A. Wang, {\it
Generalized algebra within a nonextensive statistics}, Rep. Math.
Phys. {\bf 52}, 437 (2003).

\bibitem{qborges1}E.P. Borges, {\it A q-generalization of circular and
hyperbolic functions}, J. Phys. A: Math. Gen. {\bf 31}, 5281 (1998).

\bibitem{qborges}E.P. Borges, {\it A possible deformed algebra and
calculus inspired in nonextensive thermostatistics}, Physica A {\bf
340}, 95 (2004).

\bibitem{MendesTsallis2001}R.S. Mendes and C. Tsallis, {\it Renormalization group approach to nonextensive statistical mechanics}, Phys. Lett. A {\bf 285}, 273 (2001).

\bibitem{TsallisBukman}C. Tsallis and D.J. Bukman, {\it Anomalous diffusion in the presence of external forces: exact time-dependent solutions and their thermostatistical basis}, Phys. Rev. E {\bf 54}, R2197 (1996).

\bibitem{TsallisGellmannSato}C. Tsallis, M. Gell-Mann and Y. Sato, {\it
Asymptotically scale-invariant occupancy of phase space makes the
entropy $S_q$ extensive}, Proc. Natl. Acad. Sc. USA {\bf 102}, 15377
(2005).

\bibitem{Tsallistriplet}C. Tsallis, {\it Dynamical scenario for
nonextensive statistical mechanics}, in {\it News and Expectations
in Thermostatistics}, eds. G. Kaniadakis and M. Lissia, Physica A
{\bf 340}, 1 (2004).

\bibitem{BurlagaVinas}L.F. Burlaga and A.F.-Vinas, {\it Triangle for
the entropic index $q$ of non-extensive statistical mechanics
observed by Voyager 1 in the distant heliosphere}, Physica A {\bf
356}, 375 (2005).

\bibitem{BurlagaNessAcuna}L.F. Burlaga, N.F. Ness, M.H. Acuna, {\it Magnetic fields in the heliosheath and distant heliosphere: Voyager 1 and 2 observations during 2005 and 2006}, Astrophys. J. {\bf 668}, 1246
(2007).

\end{thebibliography}
\end{document}